\documentclass[twocolumn,preprintnumbers,amsmath,amssymb]{revtex4-1}
\usepackage{graphicx}% Include figure files
\usepackage{ulem}
\usepackage[svgnames]{xcolor}
\usepackage{bm}% bold math
\usepackage{multirow}
\newif\ifshowcomments\showcommentstrue
\begin{document}

\title{Phonon thermal Hall effect in strontium titanate}

\author{Xiaokang Li$^{1,2}$, Beno\^{\i}t Fauqu\'e$^{3}$, Zengwei Zhu$^{2,}$ and Kamran Behnia$^{1,4}$}

\affiliation{(1) Laboratoire de Physique et d'Etude des Mat\'{e}riaux (CNRS)\\ ESPCI Paris, PSL Research University, 75005 Paris, France\\
(2) Wuhan National High Magnetic Field Center and
 School of Physics, Huazhong University of Science and Technology,  Wuhan  430074, China\\
(3) JEIP, USR 3573 CNRS, Coll\`ege de France, PSL University, 11, place Marcelin Berthelot, 75231 Paris Cedex 05, France\\
(4) II. Physikalisches Institut, Universit\"{a}t zu K\"{o}ln, 50937 K\"{o}ln, Germany}

\date{\today}
\begin{abstract}
 It has been known for more than a decade that phonons can produce an off-diagonal thermal conductivity in presence of magnetic field. Recent studies of thermal Hall conductivity,  $\kappa_{xy}$, in a variety of contexts, however, have assumed a negligibly small phonon contribution. We present a study of $\kappa_{xy}$ in quantum paraelectric SrTiO$_3$, which is a non-magnetic insulator and find that its peak value exceeds what has been reported in any other insulator, including those in which the signal has been qualified as 'giant'. Remarkably, $\kappa_{xy}(T)$ and $\kappa(T)$ peak at the same temperature and the former decreases faster than the latter at both sides of the peak. Interestingly, in the case of La$_2$CuO$_4$ and $\alpha$-RuCl$_3$, $\kappa_{xy}(T)$ and $\kappa(T)$ peak also at the same temperature. We also studied  KTaO$_3$  and found a  small  signal, indicating that a sizable $\kappa_{xy}(T)$ is not a generic feature of quantum paraelectrics. Combined to other observations, this points to a crucial role played by antiferrodistortive domains  in generating $\kappa_{xy}$ of this solid.
\end{abstract}
\maketitle
%\section*{Introduction}
%SrTiO$_3$ is a perovskite of the ABO$_3$ family and quantum paraelectric \cite{Müller1979}, avoiding a ferroelectric instability thanks to the quantum fluctuation. With the introduction of n-type carriers, it becomes a metal, and the most dilute superconductor after removing extremely tiny number of oxygen atoms \cite{Lin2013,Collignon2018}. SrTiO$_3$ is cubic lattice at room temperature, but it becomes tetragonal below 105K after the antiferrodistortive (AFD) transition \cite{Shirane1969}. Three possible domains will be formed in the AFD states according to the orientation of the c-axis \cite{Tao2016}. And varing faster than $T^3$ of the $\kappa$ was fund in a narrow temperature window below the peak, attributes to the Poiseuille flow \cite{Martelli2018}.
\begin{figure*}
\centering
\includegraphics[width=18cm]{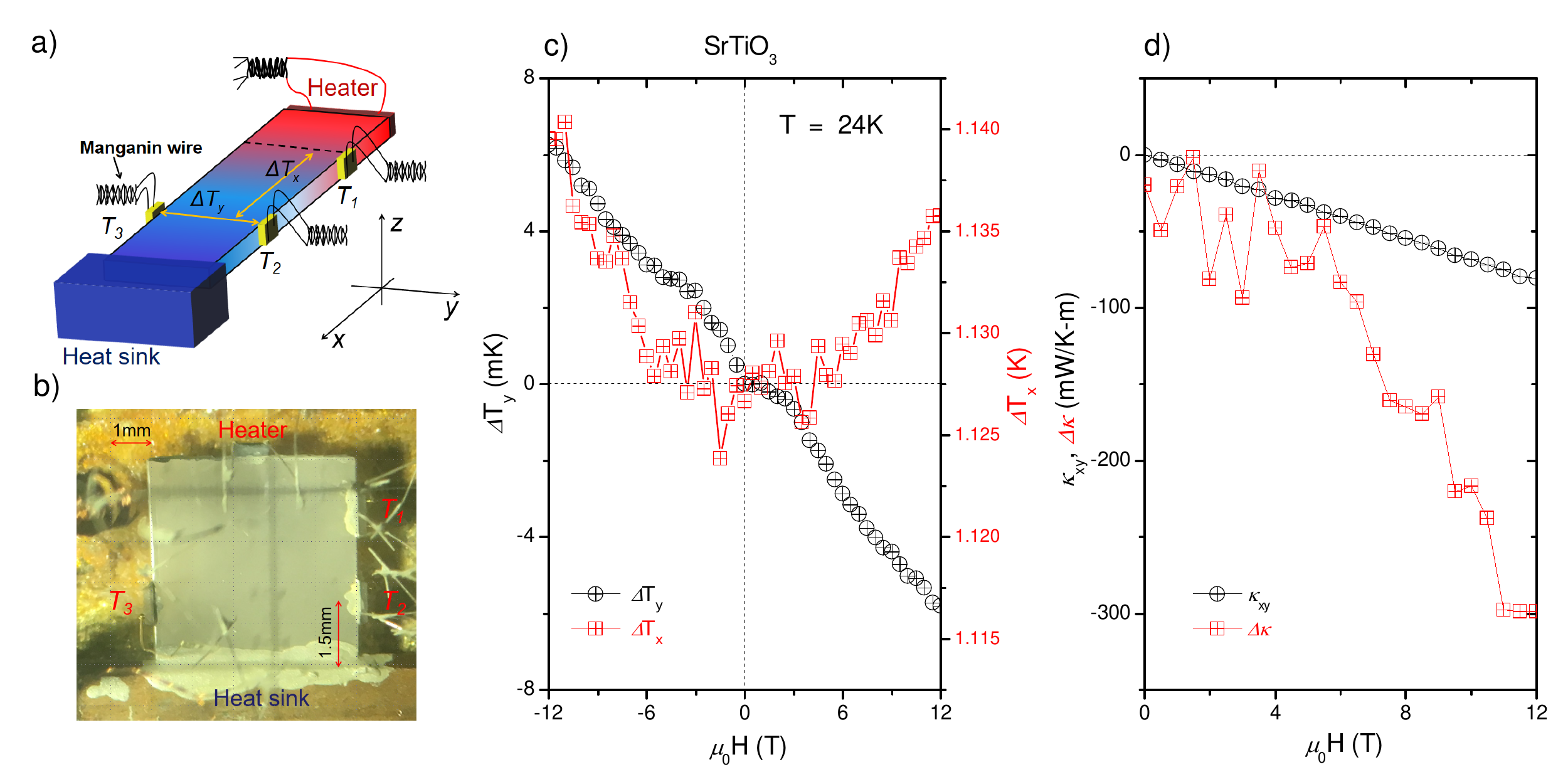}
\caption{\textbf{ Quantifying thermal Hall effect in SrTiO$_3$:} (a) Set-up for measuring longitudinal and transverse thermal differences ( $\Delta$T$_x$= T$_1$-T$_2$, $\Delta$T$_y$= T$_3$-T$_2$) generated by a longitudinal thermal current.  (b) A photograph of the sample and the set-up. The heater and the heat sink were connected to two sides of the sample and at the same level. Three thermometers were mounted near the middle of the sample.  (c) Field dependence of $\Delta$T$_y$ and $\Delta$T$_x$ at T$_2$ = 24K (labeled as sample temperature T below),  $\Delta$T$_y$ has been shifted vertically to cancel an unavoidable misalignment offset. Note that $\Delta$T$_x$ is even dominant and $\Delta$T$_y$ is odd dominant in magnetic field. (d) Extracted thermal Hall conductivity $\kappa_{xy}$ and field-induced change in thermal conductivity $\Delta\kappa = \kappa(\mu_{0}H) - \kappa(0)$ as function of field. The latter signal is noisier, because the measurement of $\Delta$T$_x$ has not been done in differential mode.}
\label{fig:Larg-THE-STO}
\end{figure*}

In most insulators, thermal conductivity can be understood with reasonable accuracy by picturing phonons as carriers of heat scattered either by other phonons or by defects and boundaries \cite{Berman}. An impressive agreement between experimental data near room temperature and \textit{ab initio} solutions of the Peierls-Boltzmann equation has been achieved in the last few years \cite{Lindsay2019}. Since phonons are neutral quasi-particles lacking magnetic moment, one may assume that their path is not affected by a magnetic field and therefore, in contrast to magnons and electrons, they cannot give rise to a transverse response.  However, experiments carried out more than a decade ago \cite{Strohm2005} showed that there is a finite measurable phonon Hall effect. The appearance of a finite transverse thermal gradient upon application of a longitudinal heat current, implied a finite $\kappa_{xy}$ in Tb$_3$Ga$_3$O$_{12}$, a paramagnetic insulator \cite{Strohm2005,Inyushkin2007}. The experimental observation motivated numerous theoretical studies \cite{Sheng2006,Kagan2008,Wang2009,Zhang2010,Agarwalla,Qin2012,Mori2014} invoking a variety of possible sources of phonon Hall effect including spin-phonon coupling \cite{Sheng2006,Kagan2008,Wang2009}, phonon Berry curvature \cite{Zhang2010,Qin2012}, skew scattering \cite{Mori2014} or simply ionic bonding \cite{Agarwalla}. 

During the past few years,  thermal Hall effect was studied in magnetic insulators \cite{Onose2010, Liu2017, Murakami2017}, spin-liquid candidates \cite{Hirschberger2015, Sugii2017} and multi-ferroics \cite{Ideue2017}. These studies of $\kappa_{xy}$ mostly assumed a marginal contribution by phonons and the detected signal was often (but not always \cite{Sugii2017}) attributed to magnetic excitations. More recently, $\kappa_{xy}$ has been measured in the Kitaev spin-liquid candidate $\alpha$-RuCl$_{3}$ \cite{Kasahara2018,Kasahara2018n,Hentrich2019} and in cuprates \cite{Grissonnanche2019}. In both cases, the observed signal was assumed to exceed significantly what could be purely a phononic contribution. 

In this paper, we present a study of thermal Hall effect in SrTiO$_3$ crystals, a quantum paraelectric \cite{Müller1979} with a variety of remarkable properties \cite{Collignon2018}.  We found a sizable $\kappa_{xy}$ in this solid. Since phonons are the unique heat carriers in this non-magnetic band insulator, it is hard to see how carriers other than phonons can cause the observed $\kappa_{xy}$. The magnitude of the observed signal is twice larger than what was reported in LaCuO$_4$ \cite{Grissonnanche2019}. However, at 15T, $\kappa_{xy}$ remains 400 times smaller than $\kappa_{xx}$ and calling this a `giant' thermal Hall effect \cite{Ideue2017,Grissonnanche2019} does not appear as an illuminating choice. 

The study of three different crystals showed that while the peak $\kappa_{xy}$ can vary  from one sample to another, the overall temperature dependence remains the same. This sample dependence is reminiscent of what was reported in $\alpha$-RuCl$_3$ \cite{Kasahara2018,Kasahara2018n,Hentrich2019}. Comparing the temperature dependence of $\kappa$ and $\kappa_{xy}$ in SrTiO$_3$, it becomes clear that they both peak at the same temperature, but the decrease in $\kappa_{xy}$ is sharper both below and above the peak temperature. We note that  in both $\alpha$-RuCl$_3$ \cite{Hentrich2019} and La$_2$CuO$_4$ \cite{Grissonnanche2019}  $\kappa$ and $\kappa_{xy}$ peak at the same temperature. We also studied KTaO$_3$, another quantum paraelectric with no antiferrodistortive transition and found that its $\kappa_{xy}$ is much smaller. This observation indicates a crucial role played by polar domain walls of SrTiO$_3$ in generating  $\kappa_{xy}$. This hypothesis is strengthened by detailed study of how the amplitude of the signal in the same sample is affected by its thermal history after trips across the 105 K structural transition. 

A member of the perovskite ABO$_3$ family, SrTiO$_3$ avoids a ferroelectric instability thanks to the quantum fluctuations \cite{Müller1979}. Upon the introduction of a tiny amount of mobile electrons, this wide-gap insulator turns to a dilute metal \cite{Spinelli} subject to a superconducting instability \cite{Schooley1964,Lin2013} and displaying non-trivial charge transport at room temperature \cite{Lin2017}. Its cubic crystal structure at room temperature is lost below 105K \cite{Shirane1969}. This structural transition is antiferrodistortive (AFD): neighboring TiO$_6$ octahedra tilt clockwise and anti-clockwise. As a consequence, the tiny tetragonal distortion generates significant  anisotropy in charge transport \cite{Tao2016}. In absence of strain, three possible domains can be present \cite{Tao2016}. The polar walls between these  micrometric domains have been a subject of numerous studies \cite{Buckley1999,Scott2012,Salje2013}.  

Previous studies of heat transport in this solid \cite{Steigmeier, Lin2014
Martelli2018} uncovered two remarkable features. In a pioneer study, Steigmeier \cite{Steigmeier} determined the temperature dependence of the thermal conductivity (which peaks to 30 W/K-m  at T$\simeq 20 K$), and found that below the maximum, it depends on the applied electric field. More recently, Martelli and co-workers \cite{Martelli2018} found that thermal conductivity decreases faster than $T^3$ below the peak. Such a behavior has only been observed in a handful of solids and attributed to the Poiseuille flow of phonons \cite{Beck}, triggered by abundant normal (i.e. momentum-conserving) collisions among phonons. Soft phonons, either those associated with the antiferrodistortive transition \cite{Yamada1969,Vogt}, or modes corresponding to the aborted ferroelectricity  \cite{Vogt}, are suspected to drive these unusual features of heat transport \cite{Martelli2018}. This may also be the case of the observation reported in the present paper.

We measured the thermal Hall effect by using a one-heater-three-thermometers method as shown in Fig.~\ref{fig:Larg-THE-STO}a,b (See the Supplemental Material \cite{supplement}, for more details).  Fig.~\ref{fig:Larg-THE-STO}c-d shows the data at 24 K. As seen in Fig.~\ref{fig:Larg-THE-STO}c, $\Delta$T$_x$ is an even and  $\Delta$T$_y$ is an odd function of magnetic field.  $\Delta$T$_y$, shifted vertically to zero by a tiny quantity due to unavoidable misalignment between $T_{2}$ and $T_{3}$,  has opposite signs for positive and negative magnetic fields, which implies $\kappa_{xy}(\mu_{0}H)=-\kappa_{xy}(-\mu_{0}H)$, as one would expect for the off-diagonal component of the conductivity tensor. On the other hand, $\Delta$T$_x$ is finite at zero field and increases symmetrically with applied magnetic field implying $\kappa(\mu_{0}H)=\kappa(-\mu_{0}H)$.
\begin{figure}
\includegraphics[width=9cm]{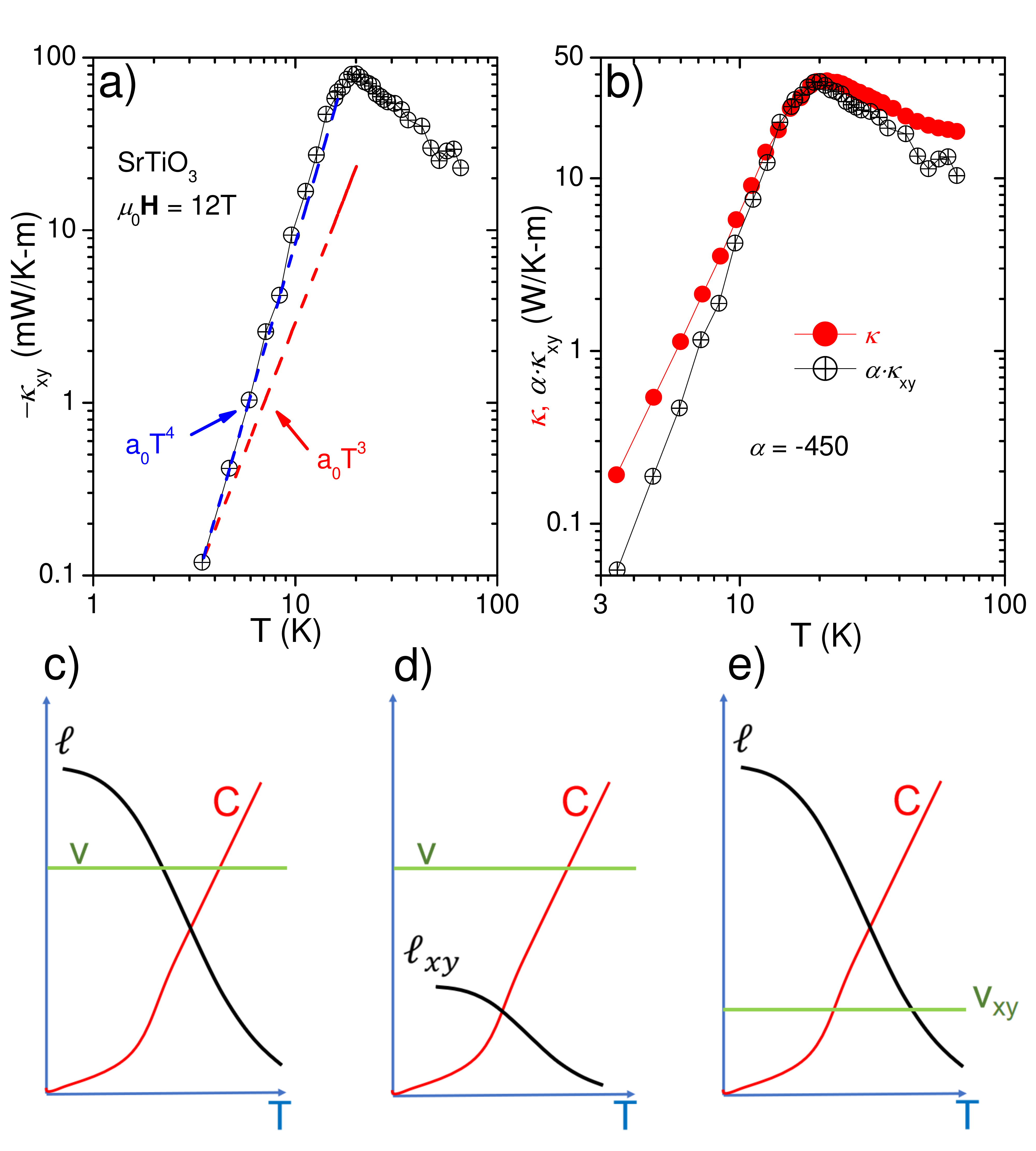}
\caption{\textbf{Thermal Hall conductivity and its correlation with longitudinal thermal conductivity :} (a) The temperature dependence of $\kappa_{xy}$ in presence of a magnetic field of 12T in SrTiO$_3$. (b) A comparison of the temperature dependence of longitudinal and transverse thermal conductivity, $\kappa_{xy}$ has been multiplied by a factor $\alpha$ equal to -450. Both peak at the same temperature, but $\kappa_{xy}$ falls faster at both sides of the peak. (c) Schematic sketch of the temperature dependence of specific heat, mean-free-path and velocity generating a peak in $\kappa$. Off-diagonal response may be caused by the skew scattering (transverse mean-free-path $\ell_{xy}$ (d) or the off-diagonal velocity v$_{xy}$ (e)).}
\label{fig:STO}
\end{figure}

The field dependence of thermal Hall conductivity $\kappa_{xy}$ and the field-induced change in thermal conductivity $\Delta\kappa = \kappa(\mu_{0}H) - \kappa(0)$ at T= 24 K are plotted in Fig.~\ref{fig:Larg-THE-STO}\color{blue}d\color{black}. The magnitude of $\kappa_{xy}$  attains -80 mW/K-m,  twice larger than the maximum observed in cuprates \cite{Grissonnanche2019}. As seen in the figure, however, this is four times smaller than the field-induced change in longitudinal thermal conductivity, $\Delta\kappa$, itself about one percent of total signal. We note that Jin \textit{et al.} \cite{Jin2015} have recently reported on a similar field-induced decrease in the lattice thermal conductivity of another band insulator.
%The thermal Hall effect result in SrTiO$_3$ was shown in Fig.~\ref{fig:Larg-THE-STO}c-f. The $\Delta$T$_y$ signal was measured two times at the T$_2$ = 24K, one with the heat current and another one without. As the Fig.~\ref{fig:Larg-THE-STO}c shows, the response with the heat current is a large and obvious asymmetric signal, however the response without is a almost flat background. After subtracted the asymmetric component of the $\Delta$T$_y$ response with the heat current in Fig.~\ref{fig:Larg-THE-STO}c, we obtained a pure linear thermal Hall effect signal, as shown in Fig.~\ref{fig:Larg-THE-STO}d, And the thermal Hall conductivity calculated using formula (2) above is around -80mW/K-m at T$_2$ = 24K. In order to vertify the authenticity of the signal, we reproduced the measurement with the different heat power but kept sample temperature T$_2$ = 24K. The result is convictive and obvious, as shown in Fig.~\ref{fig:Larg-THE-STO}e, the $\Delta$T$_y$ keeps a linear increasing and the $\kappa_{xy}$ keeps a constant with the heat power \cite{Martelli2018}. The $\kappa_{xy}$ with the temperature dependent data was shown in Fig.~\ref{fig:Larg-THE-STO}f, has a peak about -80 mW/K-m around 22K, which is twice more larger than the cuperate in recent report \cite{Grissonnanche2019}. At low temperature, the $\kappa_{xy}$ decreases so sharply, faster than cubic and even near biquadrate, and this behavior in $\kappa_{xy}$ has never been reported before.
\begin{figure}
%\centering
\includegraphics[width=9cm]{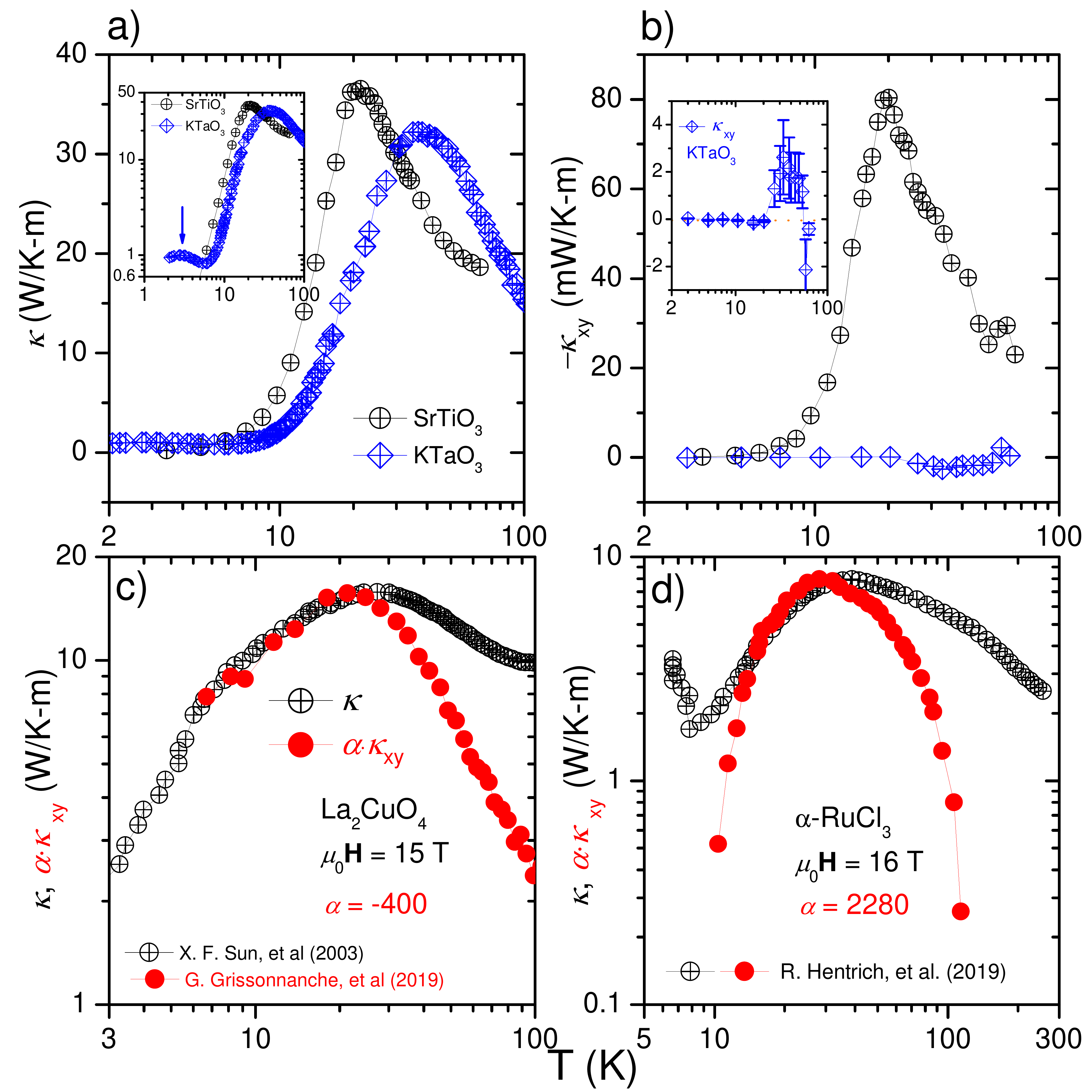}
\caption{\textbf{$\kappa_{xy}$ and $\kappa$ in other insulators:} (a) Longitudinal thermal conductivity $\kappa$ in SrTiO$_3$ and KTaO$_3$, the inset shows a logarithmic plot. (b) Transverse thermal conductivity $\kappa_{xy}$ in SrTiO$_3$ and KTaO$_3$. The inset shows a zoom showing the magnitude of the resolved signal. (c) A comparison of the temperature dependence of $\kappa_{xy}$ at 15 T (multiplied by -400) with $\kappa$ \cite{Sun2003} in La$_2$CuO$_4$. (d) The same comparison in the case of $\alpha$-RuCl$_3$. Here, $\kappa_{xy}$ at 16T is multiplied by 2280. In all these cases, $\kappa_{xy}$ and $\kappa$ peak at the same temperature and the transverse signal decreases more rapidly below and above the peak.}
\label{fig:three-Insulators}
\end{figure}

The temperature dependence of $\kappa_{xy}$ in SrTiO$_3$ is shown  in  Fig.~\ref{fig:STO}a.  As seen in the figure,  in a magnetic field of 12 T, it peaks to  -0.08 W/K-m at $\simeq 20 K$ and falls rapidly at both sides of this peak. Panel b presents a comparison of the temperature dependence of $\kappa_{xy}$ and $\kappa$, with the former multiplied by a factor $\alpha = -450$. Both peak at the same temperature, but the decrease in $\kappa_{xy}$ is faster on either sides of the maximum. As found previously \cite{Martelli2018}, $\kappa$ in SrTiO$_3$ follows T$^\beta$ (with $\beta$ slightly larger than 3) below the peak temperature. $\kappa_{xy}$ decreases even more sharply in this regime, and it almost follows a $T^{4}$ temperature dependence. With warming, the drop in the transverse signal is slightly sharper than the drop in the longitudinal one. 
\begin{figure*}
\centering
\includegraphics[width=18cm]{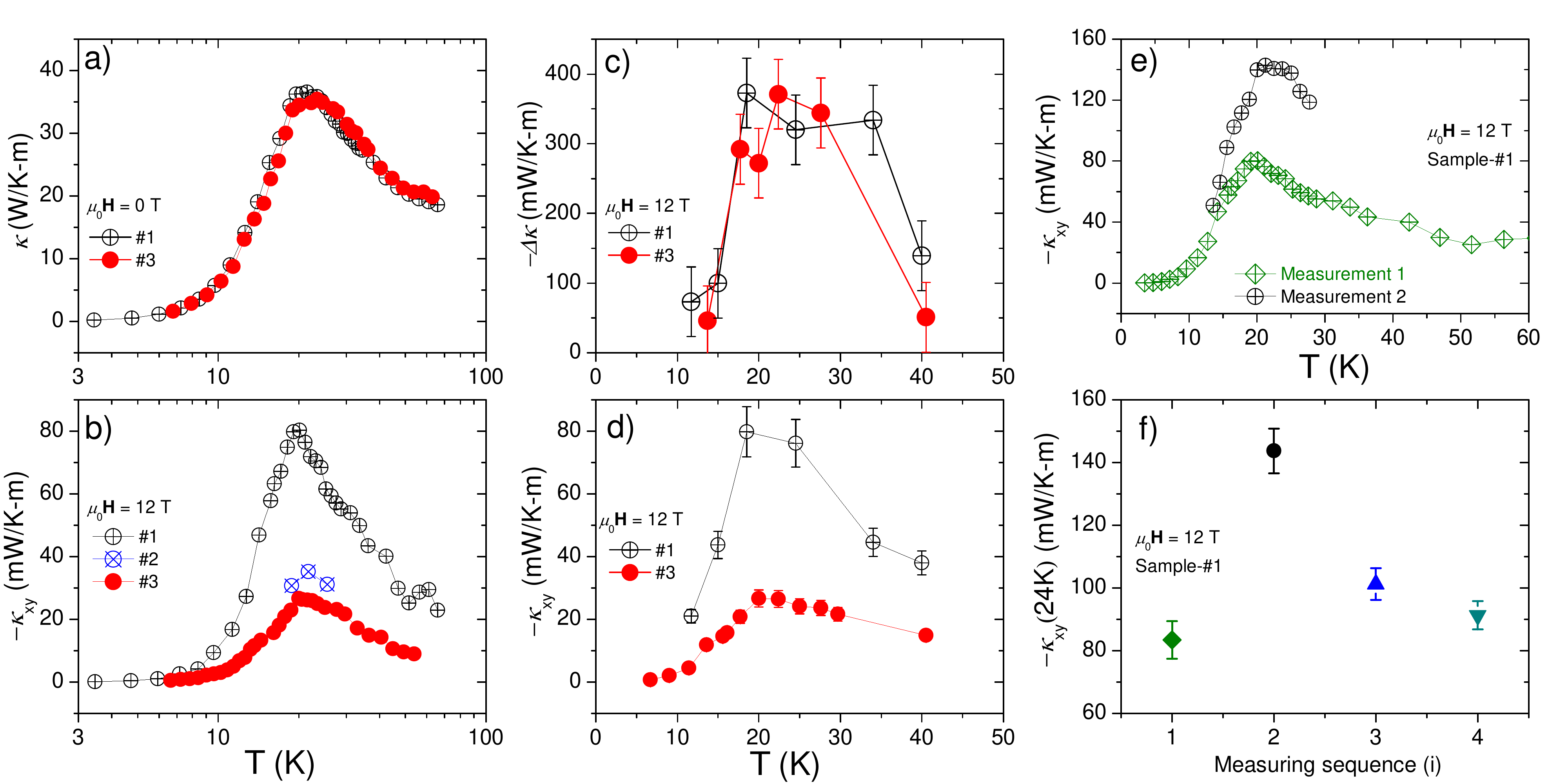}
\caption{\textbf{Variability of $\kappa_{xy}$ in SrTiO$_3$:} (a) $\kappa(T)$ in two different samples, the data is almost same. (b) Temperature dependence of $\kappa_{xy}$ in three SrTiO$_3$ samples. (c-d) Two samples with the same field-induced change in $\kappa$ display a threefold difference in their $\kappa_{xy}$, data extracted from the field-sweeping curves. (e) Even in the same sample, the magnitude of $\kappa_{xy}$ changes after warming it above T$_{AFD}$. The two curves show the data measured before (measurement 1) and after (measurement 2) being warmed to room temperature and staying in air for days. (f) $\kappa_{xy}$ of the same sample at 24K as a function of measuring sequence, separated by the warming history of being warmed to room temperature and staying in air for days, data extracted from the field-sweeping curves.}
\label{fig:Reproduced-measurement}
\end{figure*}

A phenomenological picture of $\kappa$ equates it with a product of specific heat, C, velocity, v, and mean-free-path, $\ell$. This should be summed over different phonon modes, indexed $\lambda$:

\begin{equation}\label{kappaD}
\kappa= \frac{1}{\nu}\sum_{\lambda} C^{\lambda}v^{\lambda}\ell^{\lambda}
\end{equation}

Here, $\nu$ is a dimension-dependent normalisation factor.  Usually, the variation of sound velocity with temperature is negligible. Indeed,  the experimentally measured  elastic moduli of SrTiO$_3$ \cite{Rehwald} (and therefore its sound velocity) changes by less than a few percent in our temperature range of interest. The thermal evolution of the mean-free-path and the specific heat, on the other hand, is strong and opposite to each other (see Fig.~\ref{fig:STO}c). Therefore, in an insulator $\kappa$ peaks at a temperature where the  global phonon trajectory (i.e. phonon  population times phonon mean-free-path) is maximal. This temperature has a physical significance. Thermal conductivity is most vulnerable to the introduction of point defects near this peak temperature \cite{Berman}. Our observation that $\kappa_{xy}$ peaks at this very temperature is a source of information on what causes the transverse signal. Phenomenologically, a finite $\kappa_{xy}$ implies either an off-diagonal (temperature-independent) velocity or an off-diagonal (temperature-dependent) mean-free-path. Therefore: 

\begin{equation}\label{kappaxy}
\kappa_{xy}= \frac{1}{\nu} \sum_{\lambda} C^{\lambda}(v_{xy}^{\lambda}\ell^{\lambda}+v^{\lambda}\ell_{xy}^{\lambda})
\end{equation}

Presumably, $\ell_{xy}$ and $v_{xy}$ are both much smaller than their longitudinal counterparts  as sketched in Fig.~\ref{fig:STO}d and Fig.~\ref{fig:STO}e. Therefore, the fact that $\kappa_{xy}$ peaks at the same temperature but decreases faster may be ascribed to one of the right-hand terms of equation 2 or their combination.  

The magnitude of $\kappa_{xy}$ in strontium titanate is two orders of magnitude larger than what was reported for Tb$_3$Ga$_3$O$_{12}$ \cite{Inyushkin2007}. This raises a natural question: can proximity to a ferroelectric (FE) quantum criticality \cite{Rowley} play a role in generating a large phonon thermal Hall effect? In order to answer this question, we investigated heat transport in KTaO$_3$. This insulator, like SrTiO$_3$, is close to a FE transition, but its low-temperature electric permittivity is five times smaller \cite{Lowndes1973}.

In agreement with what was reported before for KTaO$_3$ \cite{Tachibana2008} and SrTiO$_3$ \cite{Steigmeier,Martelli2018}, we found that the amplitude of the peak in longitudinal thermal conductivity is comparable (30-35 W/K-m) in the two perovskites (see Fig.~\ref{fig:three-Insulators}a). On the other hand, the amplitude of the thermal Hall conductivity is very different. In KTaO$_3$, $\kappa_{xy}$ is more than one order of magnitude smaller than in SrTiO$_3$ (Fig.~\ref{fig:three-Insulators}b). Let us note that even in the case of longitudinal thermal conductivity, there are remarkable differences between these two solids. Around T$\simeq$ 5K, thermal conductivity is sharply decreasing (displaying a faster than cubic temperature dependence) in SrTiO$_3$ but is increasing (presenting an additional bump, as seen in the inset of Fig.~\ref{fig:three-Insulators}a in KTaO$_3$. In other words, the consequences of anharmonicity for longitudinal heat transport is qualitatively different in these two apparently similar solids. Structurally, the most notable difference is the absence of the AFD transition in cubic KTaO$_3$ \cite{Barrett1968}, in contrast to its presence in SrTiO$_3$. This is our first evidence that this peculiar structural transition plays a role in setting the amplitude of $\kappa_{xy}$.
% which will let us to assume whether this special behavior is related to the perovskite ABO$_3$ lattice structure. So we chose a typical analogue KTaO$_3$ comparing to SrTiO$_3$, but the result was equally surprising.As seen in Fig.~\ref{fig:STO-KTO}, We compare the longitudinal and transverse thermal conductivity in SrTiO$_3$ and KTaO$_3$ simultaneously. The longitudinal thermal conductivity $\kappa$ is quite and comparable, both have a large peak and a large signal about 30-40 W/K-m. However, the transverse thermal conductivity $\kappa_{xy}$ is totally different, there is a largest $\kappa_{xy}$ signal in SrTiO$_3$ and comparing to any reported insulator, but a vanished signal in it's analogue KTaO$_3$. This comparison rules out the perovskite as the absolute reason, so which property inside SrTiO$_3$ let it has a so special transverse thermal conductivity behavior ?  We suggest it's the antiferrodistortive (AFD) transition, which occur in SrTiO$_3$ at 105K but not occur in KTaO$_3$, and we will discuss it with detail in later. 

The correlation between the position of peaks in longitudinal and transverse response in SrTiO$_3$ and KTaO$_3$ led us to put under scrutiny the published data in two other insulators. As seen in Fig.~\ref{fig:three-Insulators}c-d, according to the available data, there is a similar correlation between  $\kappa_{xy}(T)$ and $\kappa(T)$ in both La$_2$CuO$_4$ \cite{Sun2003,Grissonnanche2019} and in $\alpha$-RuCl$_3$ \cite{Hentrich2019}. In all cases, $\kappa_{xy}$ and $\kappa$ peak at almost the same temperature and the decrease in $\kappa_{xy}$ is sharper (or in one case almost equal) to the decrease in $\kappa$. We notice that this correlation, which was not reported before, indicates a major role played by the principal heat carriers in setting the transverse response. 

Our additional measurements build up the case for a prominent role played by AFD domains. The results are shown in Fig.~\ref{fig:Reproduced-measurement}. First of all, we studied three different SrTiO$_3$ samples, provided by two different companies. As shown in Fig.~\ref{fig:Reproduced-measurement}a-b, all three samples show a sizable $\kappa_{xy}$, but different amplitudes. Two samples in which the magnitude of $\kappa$ is almost the same (panel a), display a threefold difference in their peak of $\kappa_{xy}$ (panel b, d). As seen in panel (c), the field-induced decrease in $\kappa$ is roughly the same in the two samples.

In a second set of measurements, we repeated our measurements of $\kappa_{xy}$ on the same sample after warming it above T$_{AFD}=105$ K and cooling it back again. As seen in Fig.~\ref{fig:Reproduced-measurement}e-f (for more details, see the Supplemental Materials \cite{supplement}), warming above the AFD transition temperature can change the magnitude of $\kappa_{xy}$ (T=24 K) in the same sample. 

Buckley \textit{et al.} \cite{Buckley1999} observed needle-like structural domains below 105 K in SrTiO$_3$ and  ``found almost no memory of the domain patterns under repeated heating and cooling through the transition point''\cite{Buckley1999}. The typical size of the observed domains was a micron,  comparable to the apparent phonon mean-free-path extracted from longitudinal thermal conductivity and specific heat \cite{Martelli2018}. An intimate link between domain configuration and the amplitude of $\kappa_{xy}$ would explain why the amplitude of $\kappa_{xy}$ can be different after thermal cycling above T$_{AFD}$, wiping out the previous configuration of domains. Obviously, the sample dependence of the signal and its virtual absence in KTaO$_3$ also find natural explanations.

Theoretical scenarios for phonon thermal Hall effect \cite{Sheng2006,Kagan2008,Wang2009,Zhang2010,Agarwalla,Qin2012,Mori2014} either invoke skew scattering of heat carriers or let the magnetic field  generate a transverse velocity. Let us have a look to our results in either of these schemes. One may be tempted to attribute the observed $\kappa_{xy}$ to skew scattering of phonons by the AFD domain walls, which according to a number of experiments \cite{Scott2012,Salje2013} are polar. However, the skew-scattering picture would have a hard time to explain the disconnection between the field-induced decrease in $\kappa$ and the finite $\kappa_{xy}$ (Fig.~\ref{fig:Reproduced-measurement}c, d). Alternatively, one may point to the fact that the slight tetragonal distortion leads to quasi-degenerate acoustic phonon modes and it has been suggested \cite{Bussmann} that the acoustic phonons hybridize with the transverse optical phonons. Thanks to these features, the magnetic field may become able to couple to titanium-oxygen ionic bonds \cite{Agarwalla} and  generate a transverse velocity. Presumably, this should crucially depend on the relative orientation of the magnetic field and each of the three tetragonal domains; hence a dependence on precise domain configuration.  Note that \textit{Ab initio} theoretical calculations find imaginary frequencies \cite {Ashauer} for strontium titanate. Two recent theoretical studies succeeded in finding real phonon frequencies \cite {Feng,Tadano}. However, the focus of both was the cubic state and the phonon spectrum below the AFD transition remains unknown. Future theoretical studies may fill this void. Future experiments may use strain to control the configuration of domains. 

In summary, phonons in SrTiO$_3$ can generate a $\kappa_{xy}$ larger than what was reported in any other insulator. This is not generic to all quantum paraelectric solids and appear to be intimately linked to the occurrence of antiferrodistortive (AFD) transition in SrTiO$_3$. We find that not only in SrTiO$_3$, but also in other insulators $\kappa_{xy}$ and $\kappa$ peak at the same temperature. The observation appears as a clue to identify carriers and collisions which generate the transverse signal. In the case of SrTiO$_3$, two experimental observations point to the role of tetragonal domains in generating the signal.

We are grateful to Jing-yuan Chen, Ga\"el Grissonnanche, Steve Kivelson and Louis Taillefer for stimulating discussions. This work was supported by the Agence Nationale de la Recherche (ANR-18-CE92-0020-01). B. F. acknowledges support from Jeunes Equipes de l'Institut de Physique du Collge de France (JEIP). Z. Z. acknowledges support from the National Science Foundation of China (Grant No. 11574097 and No. 51861135104) and The National Key Research and Development Program of China (Grant No.2016YFA0401704). X. L. acknowledges a PhD scholarship by the China Scholarship Council (CSC).

\clearpage
% Add 'S' to the numbering inside the supplement
\renewcommand{\thesection}{S\arabic{section}}
\renewcommand{\thetable}{S\arabic{table}}
\renewcommand{\thefigure}{S\arabic{figure}}
\renewcommand{\theequation}{S\arabic{equation}}
\setcounter{section}{0}
\setcounter{figure}{0}
\setcounter{table}{0}
\setcounter{equation}{0}
{\large\bf Supplemental Material for ``Phonon thermal Hall effect in strontium titanate"}
%{\large\bf by X. Li et al.}
\setcounter{figure}{0}
\section{Samples and measurement technique} Single crystals of SrTiO$_3$ and KTaO$_3$ were obtained commercially. The dimensions of the SrTiO$_3$ samples (\#1 and \#2 from SurfaceNet, \#3 from CrysTech) were $5mm \times 5mm \times 0.5mm$ and their orientation along (100). The KTaO$_3$ sample had a dimension of $10mm \times 6.5mm \times 0.5mm$. Three thermometers (Cernox-1030) were used to simultaneously measure the longitudinal and transverse thermal gradient. 
A one-heater-three-thermometers technique was employed. The heat flow was applied along the x-axis and the magnetic field oriented along the z-axis. Then the longitudinal ($\Delta$T$_x$= T$_1$-T$_2$) and the transverse ($\Delta$T$_y$= T$_3$-T$_2$) temperature difference were detected simultaneously using three resistive chips (Cernox-1030), measuring T$_1$, T$_2$ and T$_3$. The longitudinal ($\kappa$) and the transverse ($\kappa_{xy}$) thermal conductivity were obtained using:
\begin{equation}\label{S1}
\kappa=\frac{Q}{(\Delta T_x /l)(w \cdot t)} ,
\end{equation}
\begin{equation}\label{S2}
\kappa_{xy}=\kappa \cdot \frac{\Delta T_y /w}{\Delta T_x /l} .
\end{equation}
Here, $Q$, $l$, $w$, $t$ are the heat power, length between T$_1$ and T$_2$, the sample width and thickness respectively. All the measurements were performed in a PPMS (Physical Property Measurement System) within a stable high-vacuum sample chamber. The thermal gradient in the sample was produced through a 1k$\Omega$ chip resistor alimented by a DC current source (Keithley 6220). The DC voltage on the heater was measured through a Keithley 2000. Three lock-in amplifiers (SR830) were used to monitor the T$_1$, T$_2$ and $\Delta$T$_y$ in AC mode at low frequency (f$\approx$ 17Hz). In order to attain a higher resolution, $\Delta T_y$ was measured in a differential mode: the difference between the T$_2$ and T$_3$ resistances was measured by a lock-in amplifier after being amplified by 100. Both field-sweeping and temperature-sweeping protocols were used in measuring $\kappa_{xy}$ and the two sets of data were found to match.
%A Keithley 6220 provided injected current and a Keithley 2000 measured the produced voltage of a 1k$\Omega$ chip resistor, which was used as heater to establish temperature gradient along the sample. Three lock-in amplifiers (SR830) were used to measure the signal of T$_1$, T$_2$ and $\Delta{T}_{y}$ by using a AC mode. Two transverse thermometers (T$_2$, T$_3$) with very close coefficient were chosen for measuring $\Delta{T}_{y}$ by using a differential amplification technique, the signal of T$_2$ and T$_3$ was amplified 100 times first and then the difference was measured by a lock-in amplifier. Both field-sweeping and temperature-sweeping protocols were used in measuring $\kappa_{xy}$ and the data match well.
\section{Suppression of the background signal} 
\begin{figure}
\includegraphics[width=8cm]{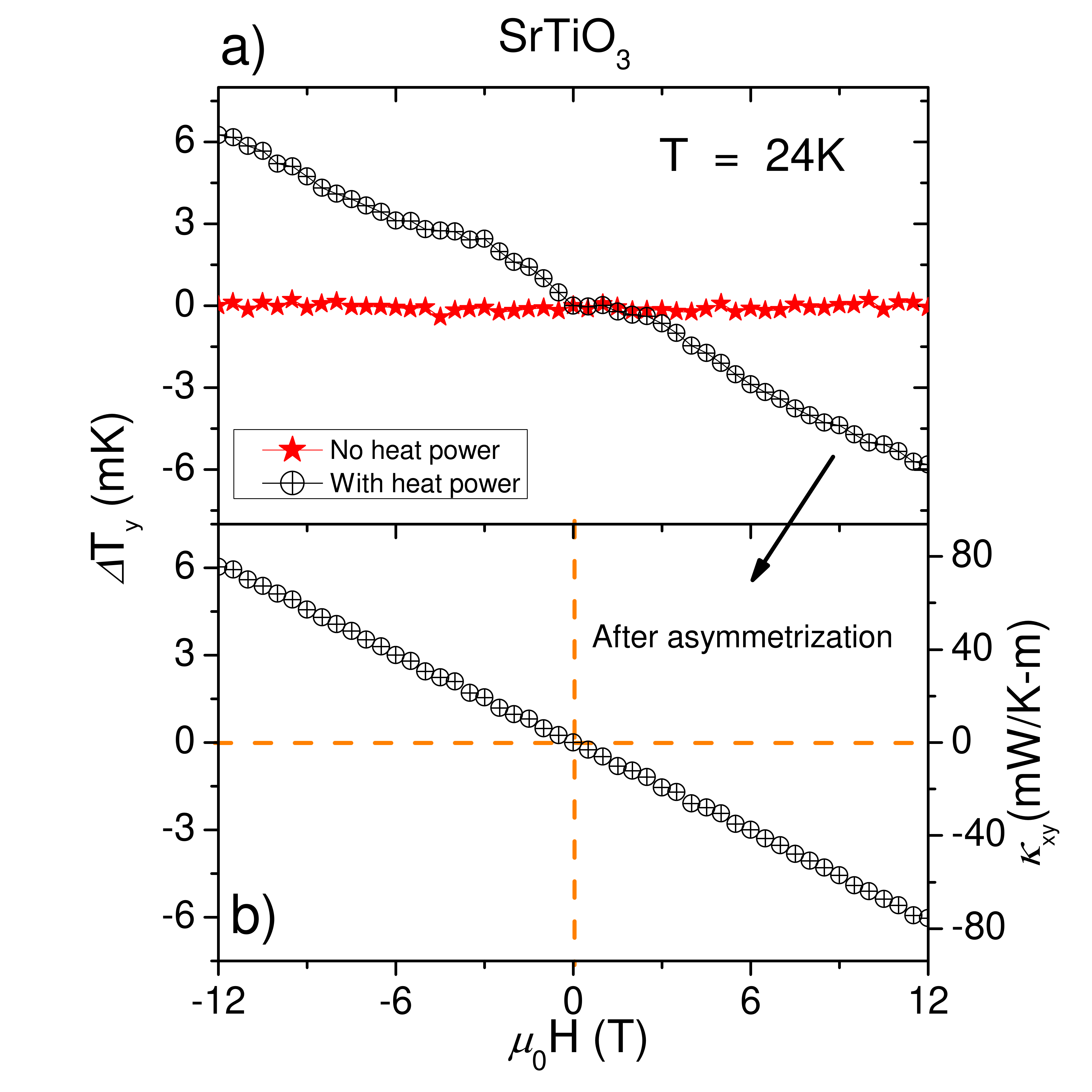}
\caption{(a) $\Delta{T}_{y}$ as a function of the magnetic field with (red star) and without (black circle) the heat power at 24K. (b) $\Delta{T}_{y}$ (left axis) and $\kappa_{xy}$ (right axis) signal after the asymmetrization.}
\label{Fig:Asy-Metho}
\end{figure}

As seen in Fig.~\ref{Fig:Asy-Metho}, there is a small change in the transverse thermal gradient ($\Delta T_y$) induced by the magnetic field even in absence of an applied heat. This background signal has two origins. First of all, the resistive chips have a finite magnetoresistance. This can be reduced by choosing two very similar thermometers. The other source of this background is the misalignment between transverse electrodes. Both effects (magneto-resistance and misalignment) are expected to be symmetric with the magnetic field, in contrast to the intrinsic thermal Hall effect signal. As illustrated in Fig.~\ref{Fig:Asy-Metho}a (black circles), the measured signal is dominantly odd. To extract the purely asymmetric component of the raw data, we anti-symmetrized it: ($\Delta{T}_{y}$ = ($\Delta{T}_{y}(+\mu_{0}H) - \Delta{T}_{y}(-\mu_{0}H))/2$). The extracted $\kappa_{xy}$ signal is linear with the magnetic field as shown in Fig.~\ref{Fig:Asy-Metho}b.  \begin{figure*}
\centering
\includegraphics[width=16cm]{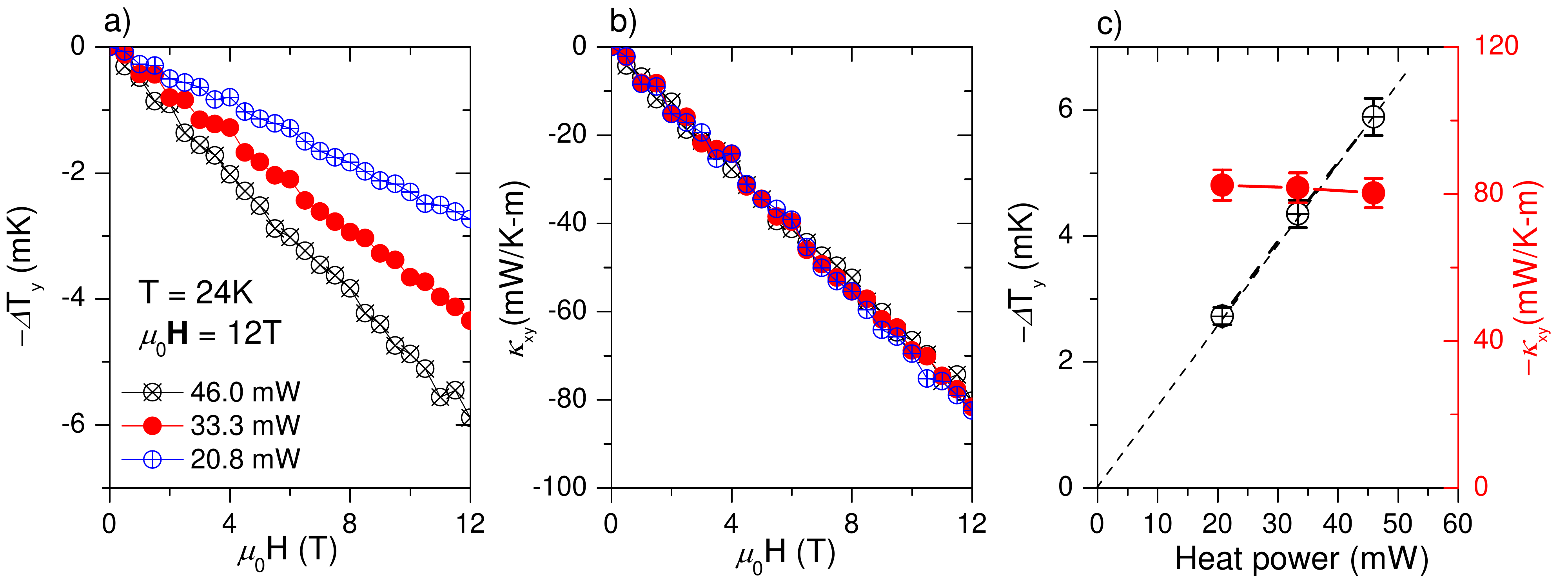}
\caption{(a-b) The field dependence of the $\Delta{T}_{y}$ and $\kappa_{xy}$ at 24K with three different heat power ($Q$=20.8mW, 33.3mW and 46.0mW). (c) $\Delta{T}_{y}$ and $\kappa_{xy}$ at 24K, $\mu_{0}H$ = 12T as a function of Q. As we expected that $\Delta{T}_{y}$ scales linearly with the heat power.}
\label{Fig:heat-power}
\end{figure*}
\begin{figure}
\includegraphics[width=8cm]{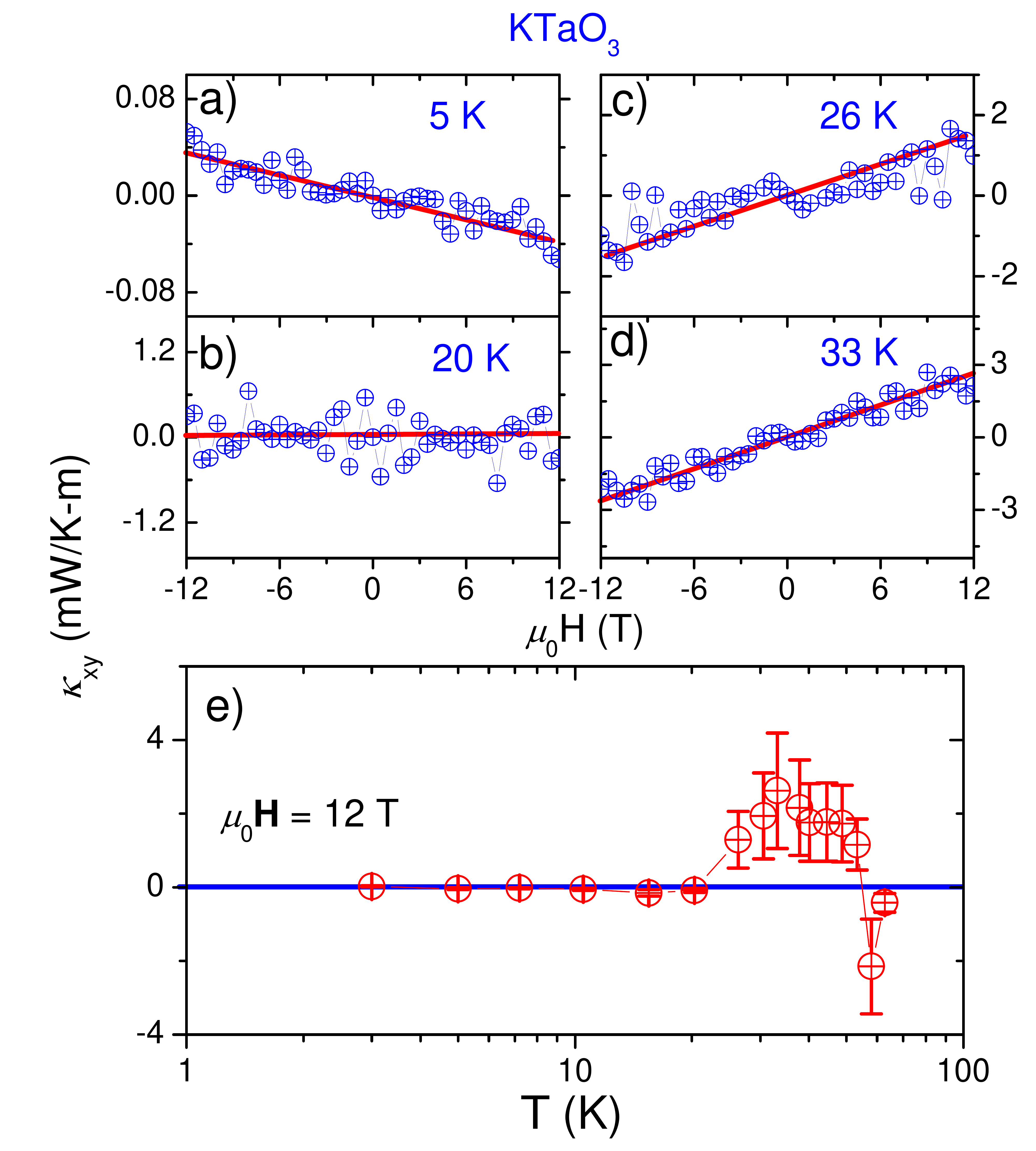}
\caption{(a-d) The field dependence of $\kappa_{xy}$ in KTaO$_3$. (e) The temperature dependence of $\kappa_{xy}$ in KTaO$_3$ extracted from the field-sweeping curves.}
\label{Fig:KTO}
\end{figure}
\section{Dependence on heat power} Combining equations S1 and S2, we get:
\begin{equation}\label{S3}
\kappa_{xy}=\kappa^{2} \cdot \frac{\Delta T_y \cdot t }{Q} .
\end{equation}
\begin{equation}\label{S4}
Q=\kappa^{2} \cdot \frac{\Delta T_y \cdot t }{\kappa_{xy}} .
\end{equation}
According to the formula S4, the transverse temperature difference $\Delta{T}_{y}$ is expected to be linear with the heat power. Fig.~\ref{Fig:heat-power}a-b presents data obtained with three different heat powers ($Q$=20.8 mW, 33.3 mW amd 46.0 mW) applied to the same sample kept at the temperature of 24K. As expected, $\kappa_{xy}$ is the same and $\Delta T_y$ is a linear function of the heat power  (Fig.~\ref{Fig:heat-power}c). \begin{figure}
\includegraphics[width=8cm]{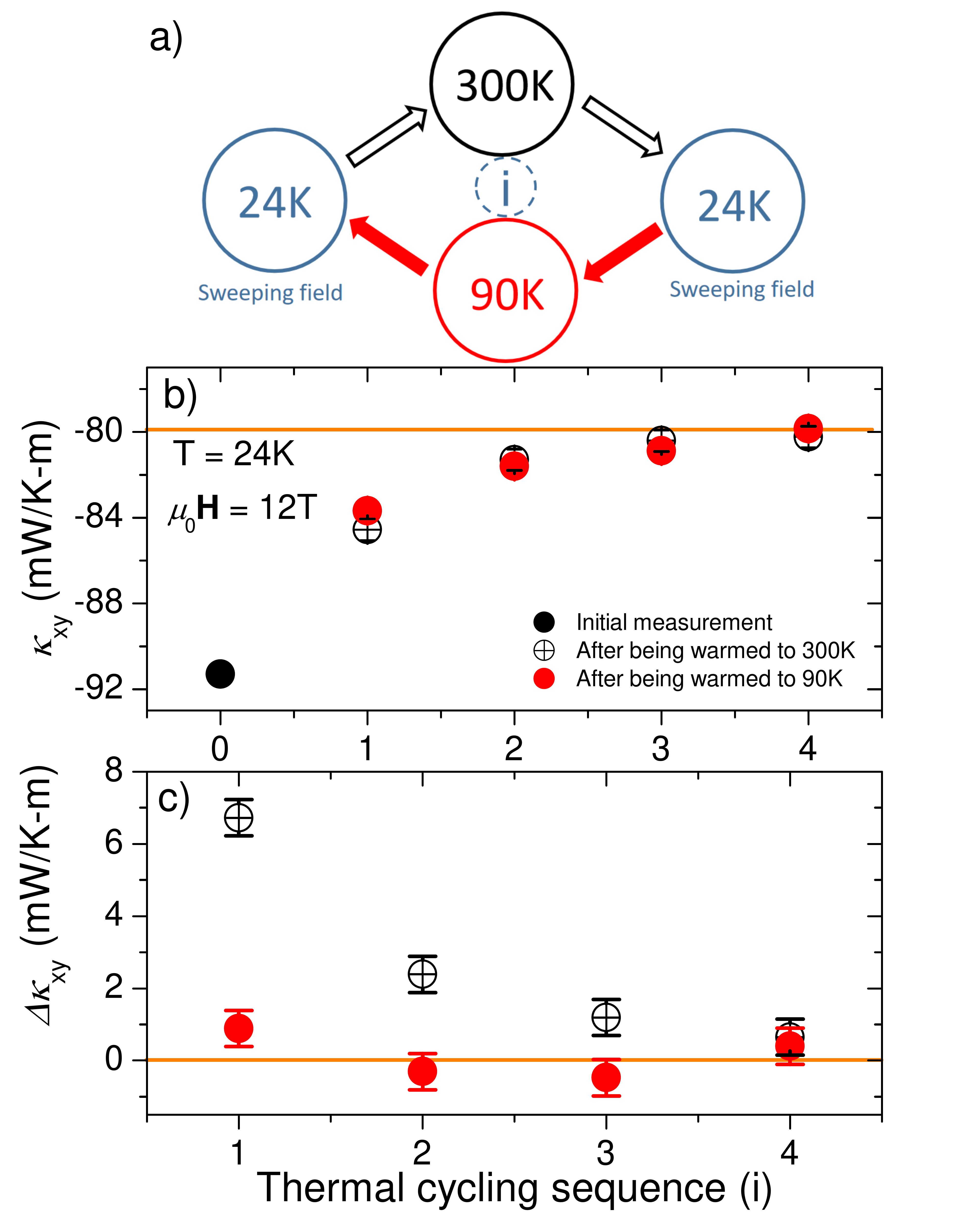}
\caption{(a) The sketch of the thermal cycling. (b-c) The $\kappa_{xy}$ and $\Delta\kappa_{xy}$ (the difference between two successive measurements at 24K) of SrTiO$_3$ in each thermal cycling sequence. }
\label{Fig:thermal-history}
\end{figure}
\section{Thermal Hall conductivity of KTO} As discussed in the main text, we observed a very small $\kappa_{xy}$ signal in KTaO$_3$, about 30 times smaller than in SrTiO$_3$. $\kappa_{xy}$ of KTaO$_3$ has a peak about 3 mW/K-m, which is 1/12000 of $\kappa$. %\sout{A wider sample and a larger heat power was used to detect this small signal.}  
We swept the field from +12T to -12T at different temperatures. In Fig.~\ref{Fig:KTO}a-d, four typical curves are shown. After extracting the $\kappa_{xy}$ value at 12T, we determined the temperature dependence of $\kappa_{xy}$, shown in Fig.~\ref{Fig:KTO}e. We note that $\kappa_{xy}$ of KTaO$_3$ is positive at its peak value, in contrast to SrTiO$_3$ and La$_2$CuO$_4$. 
Error bars in Fig.~\ref{Fig:KTO}e  are large. This is because, we needed to apply a quite large heat power (50-80mW above 20K) to detect a $\kappa_{xy}$ of such a small magnitude in KTaO$_3$. This large heat power will generate a large thermal gradient (1K/mm) and difference (3K between $T_{1}$ and $T_{2}$) on the sample, which will introduce a large error. To ensure the reliability of the data, we estimate this error large as 50$\%$. This is added to the error from  scattered data points in $\kappa_{xy}(B)$.
\section{Thermal history measurements} We reported four sequences of thermal measurement in the main manuscript (Fig. 4f). Each sequence of measurement was separated from the previous one by a warming process to room temperature and staying in air for several days. What we report here is a second set of thermal history measurements, in which each thermal cycle has two different warming temperature points, 300K and 90K, above and below the AFD transition temperature (105K) respectively. After an initial measurement of $\kappa_{xy}$ by sweeping field from +12T to -12T at 24K (labelled as thermal circling sequence 0), two other field sweeps and  measurements were performed at 24K. In each  case the sample was warmed to 300K and to 90K and was kept in high-vacuum for hours (see Fig.~\ref{Fig:thermal-history}a). Such thermal circles were repeated four times and the results are shown in Fig.~\ref{Fig:thermal-history}b-c. It's obvious that  warming to 300K can significantly change the $\kappa_{xy}(T=24K)$, but not warming to 90K . This observation points out to the role of AFD domains in setting the magnitude of $\kappa_{xy}$. Interestingly, after three cycles, $\kappa_{xy}$ is gradually stabilized and no more affected by the thermal cycling.\\ \begin{figure}
\includegraphics[width=8cm]{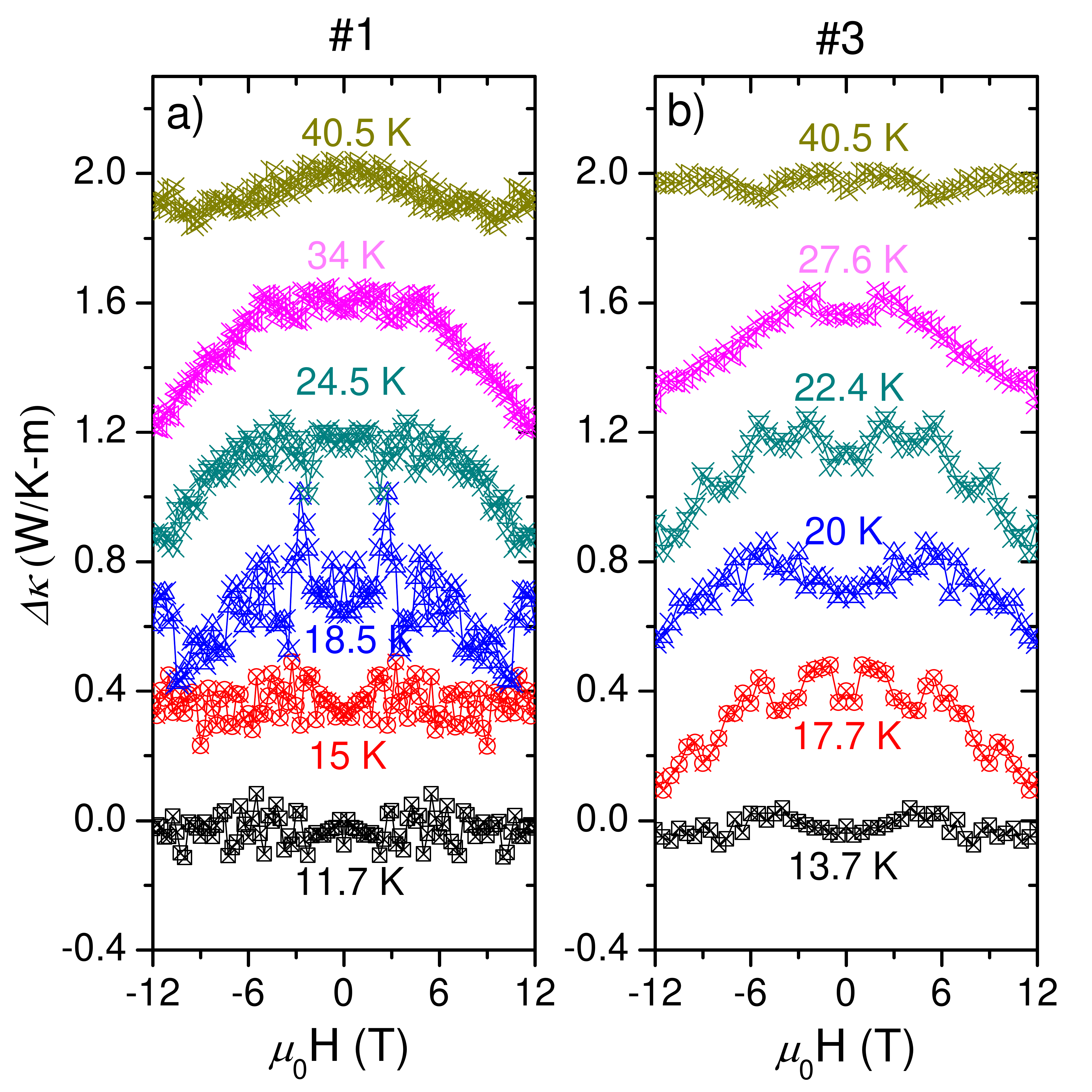}
\caption{The field dependence of the thermal conductivity in two SrTiO$_3$ samples, $\Delta\kappa$ = $\kappa(\mu_{0}H)$ - $\kappa(0)$.}
\label{Fig:Thermal-conductivity with field}
\end{figure}
\section{Field dependent thermal conductivity in STO} We also studied the The field dependence of the thermal conductivity  in two SrTiO$_3$ samples. The results,  shown in Fig.~\ref{Fig:Thermal-conductivity with field},  are roughly similar in the two samples.  In the $20 K < T< 30 K $ temperature range, a magnetic  field of 12 T induces a change ($\Delta\kappa$), which is about one percent of total $\kappa$. The magnitude of   $\Delta\kappa$ is larger than the off-diagonal response.

\end{document}